\documentclass[twocolumn,english,aps,prl,showpacs]{revtex4}
\usepackage[T1]{fontenc}
\usepackage{amsmath,graphicx,amssymb,epsfig,babel,dsfont}

\begin{document}

\title{Switchable thermal antenna by phase transition}

\author{Philippe Ben-Abdallah}
\email{pba@institutoptique.fr}
\affiliation{Laboratoire Charles Fabry,UMR 8501, Institut d'Optique, CNRS, Universit\'{e} Paris-Sud 11,
2, Avenue Augustin Fresnel, 91127 Palaiseau Cedex, France.}
\author{Henri Benisty}
\affiliation{Laboratoire Charles Fabry,UMR 8501, Institut d'Optique, CNRS, Universit\'{e} Paris-Sud 11,
2, Avenue Augustin Fresnel, 91127 Palaiseau Cedex, France.}
\author{Mondher Besbes}
\affiliation{Laboratoire Charles Fabry,UMR 8501, Institut d'Optique, CNRS, Universit\'{e} Paris-Sud 11,
2, Avenue Augustin Fresnel, 91127 Palaiseau Cedex, France.}

\date{\today}


\begin{abstract}
We introduce a thermal antenna which can be actively switched by phase transition. The source makes use of periodically patterned vanadium dioxide, a metal-insulator phase transition material which supports a surface phonon-polariton (SPP) in the infrared range in its crystalline phase. Using electrodes properly registred with respect to the pattern, the phase transition of VO2 can be localy triggered within few microseconds and the SPP can be diffracted making the thermal emission highly directionnal. This switchable antenna could find broad applications in the domain of active thermal coatings or in those of infrared spectroscopy and sensing.
\end{abstract}

\maketitle

A thermal antenna is a spatially coherent source that radiates in the infrared or the mid-infrared frequency range. In 1988, after a pionnering work of Hesketh  \cite{Hesketh86} on the radiant modes of periodic micromachined silicon surfaces, the first thermal antenna has been designed using deep gratings \cite{Hesketh88}. Since then, numerous directional thermal sources and partially coherent sources have been proposed by texturing materials at subwavelength scale  \cite{Kreiter,Greffet,Kollyukh,PBA_JOSA,Celanovic,Lee,Drevillon,Battula,Joulain,Zhang,Wang}. However, to date, no antenna can be actively controlled without moving parts and their emission patterns remain timely-invariant. An important step toward such a control  has nevertheless been recently achieved. specifically, a new strategy was proposed  to timely modulate the near-field heat flux exchanged between two hot bodies \cite{van Zwol1,van Zwol2} by using insulator-metal transition (IMT) coatings deposited on the exchange surfaces of these bodies. By passing an IMT material from its crystalline phase to its amorphous state, its optical properties can be very significantly modified. This bifurcation in the optical behavior of IMTs materials has been recently exploited to achieve efficient radiative thermal rectifiers \cite{PBA_APL}.  
In the present Letter, we introduce a switchable thermal antenna by exploiting these phase transitions. It is basically composed of an IMT material that supports a surface wave in the frequency range of emission (Planck's window) of the antenna.  By an appropriate addressing of regions where transition is triggered,  a grating can be generated on the crystalline phase of IMT material so that the surface wave can be diffracted, on the demand,  into the far field, making the emission pattern highly directional. Also, we show that the direction of emission can be controlled and the thermal emission itself can be switched off or switched on by making appropriate local phase transitions.     
\begin{figure}[Hhbt]
\includegraphics[scale=0.4]{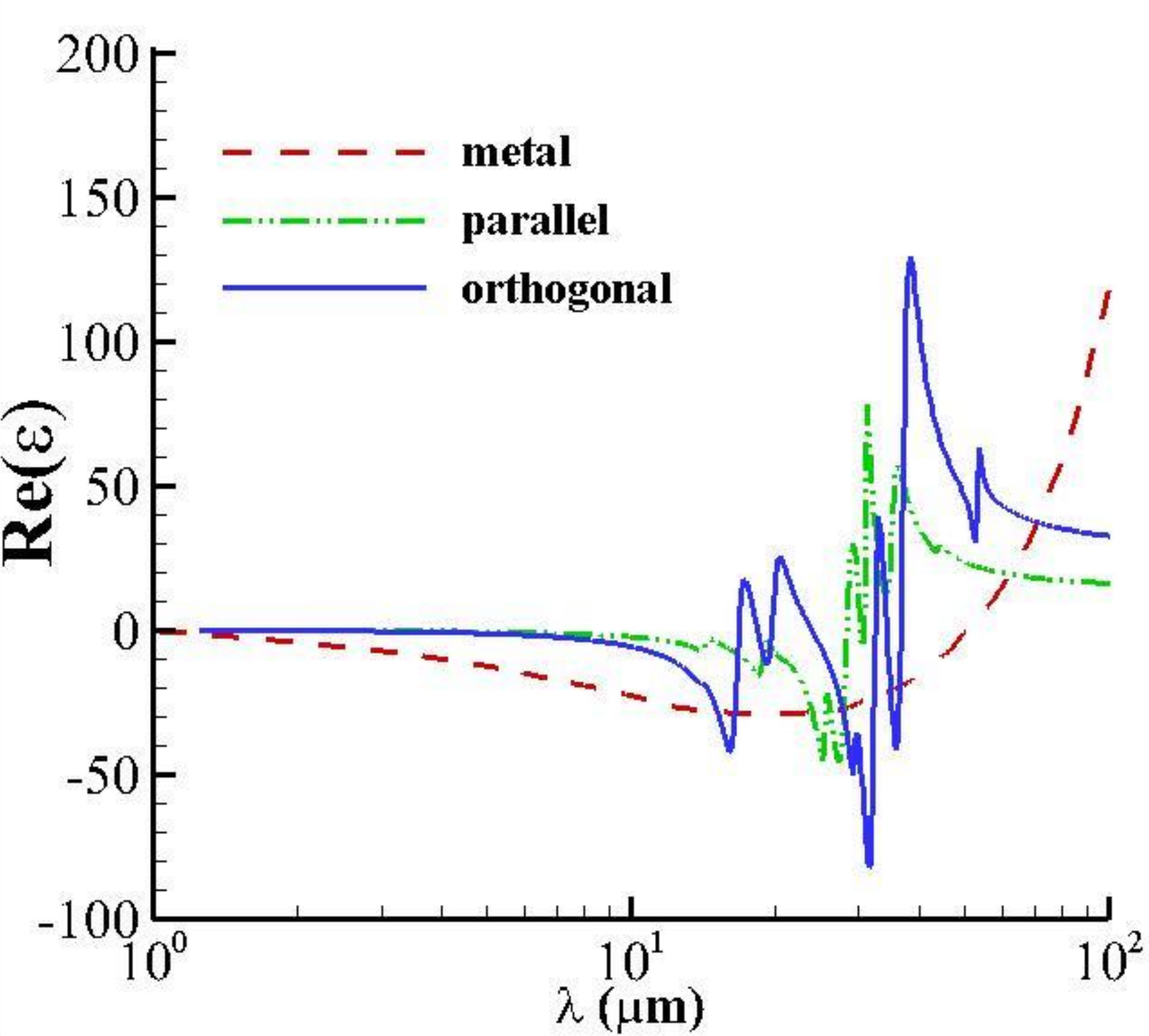}
\includegraphics[scale=0.4]{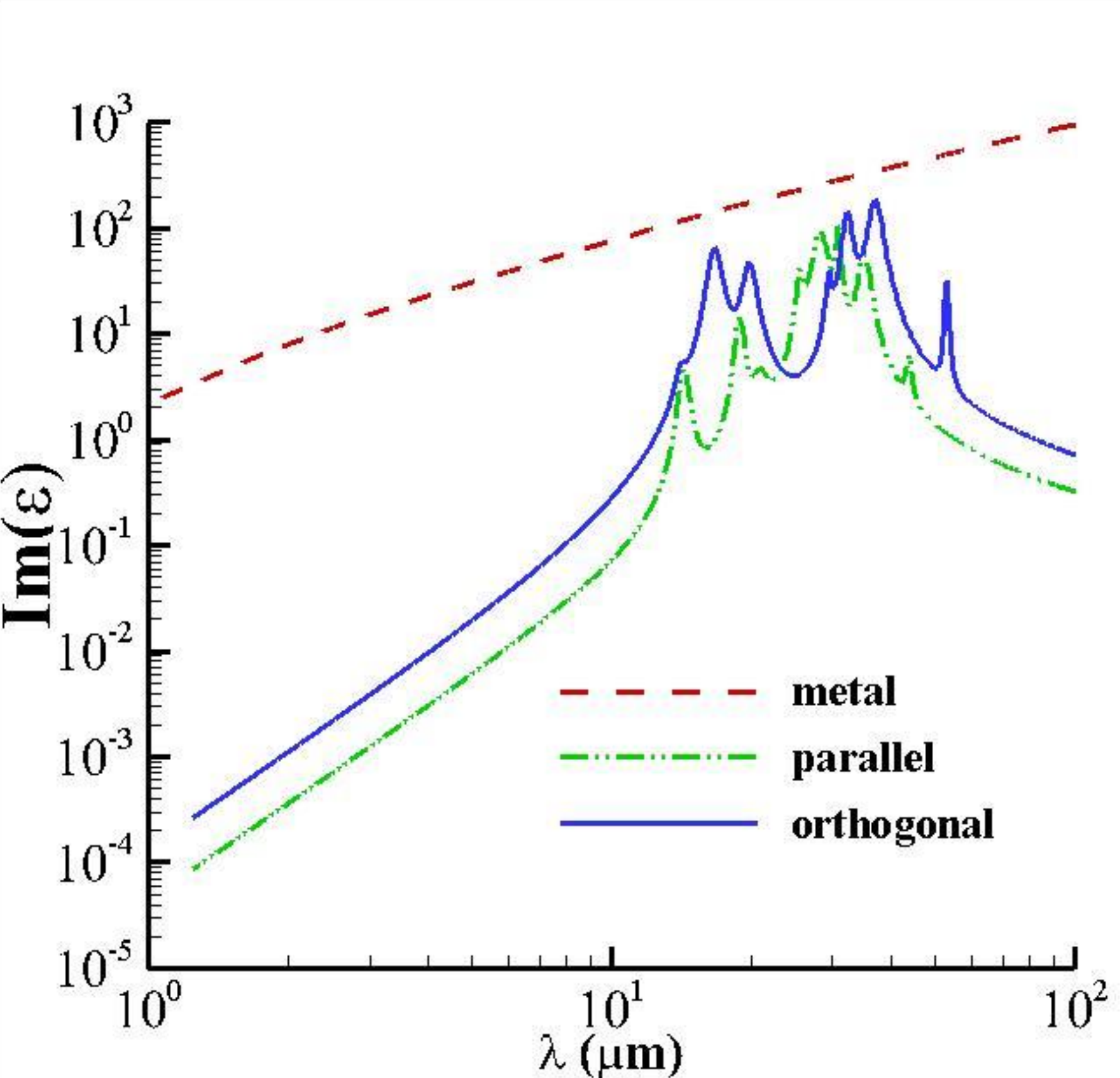}
\caption{ (a) Real parts of the permittivity $\epsilon_m$ of VO$_2$ in the metallic phase and of the permittivity $\epsilon_\parallel$ in the crystalline state along the plane  and $\epsilon_\parallel$ and $\epsilon_\perp$ along the optical axis. (b) Imaginary parts. \label{Fig:Perm}}
\end{figure}
To start, let us consider a semi-infinite sample of vanadium dioxide (VO2), an IMT material of critical temperature $T_c=340 K$ surrounded by vacuum.  Below $T_c$, this material is in its crystalline (monoclinic) phase.  In this phase, it is an uniaxial medium. Experimental datas show that \cite{Baker} its optical axis is, most of the time, orthogonal to the surface. The dielectric permittivity along this axis $\epsilon_\perp$  and in the plane parallel to the surface $\epsilon_\parallel$ are plotted in Fig. 1. In this phase VO2 supports a surface phonon-polariton in polarization TM. The dispersion relation of this surface wave plotted in Fig. 2 can be classicaly found by extracting the pole of the corresponding reflection coefficient \cite{BiehsEtAl2011}
\begin{equation}
  r^{{\rm p}}_l=\frac{\epsilon_\parallel k_{z0}-k_{p}}{\epsilon_\parallel k_{z0}+k_{p}},
  \label{Eq:r_p}
\end{equation}
where $k_{p}$ is given by the following relation
\begin{equation}
\epsilon_\parallel \epsilon_{\perp}\frac{\omega^2}{c^2}-\epsilon_\parallel \kappa^2-\epsilon_{\perp}k_{p}^2 =0
\label{Eq:Fresnel}
\end{equation}
and $k_{z0} = \sqrt{\omega^2/c^2 - \kappa^2}$, $\kappa$ being the parallel wavector component. 
Beyond $T_c$, VO2 transits in its amorphous state and becomes optically isotropic. In this phase it behaves as a metal (see Fig. 1) and it does not support anymore surface wave (Fig. 2).  Now we assume that  some regions of VO2  can be commutated from the crystalline state to the amophous one using, for instance, a network of electrodes periodically distributed as illustrated in Fig. 3. 

\begin{figure}[Hhbt]
\includegraphics[scale=0.40]{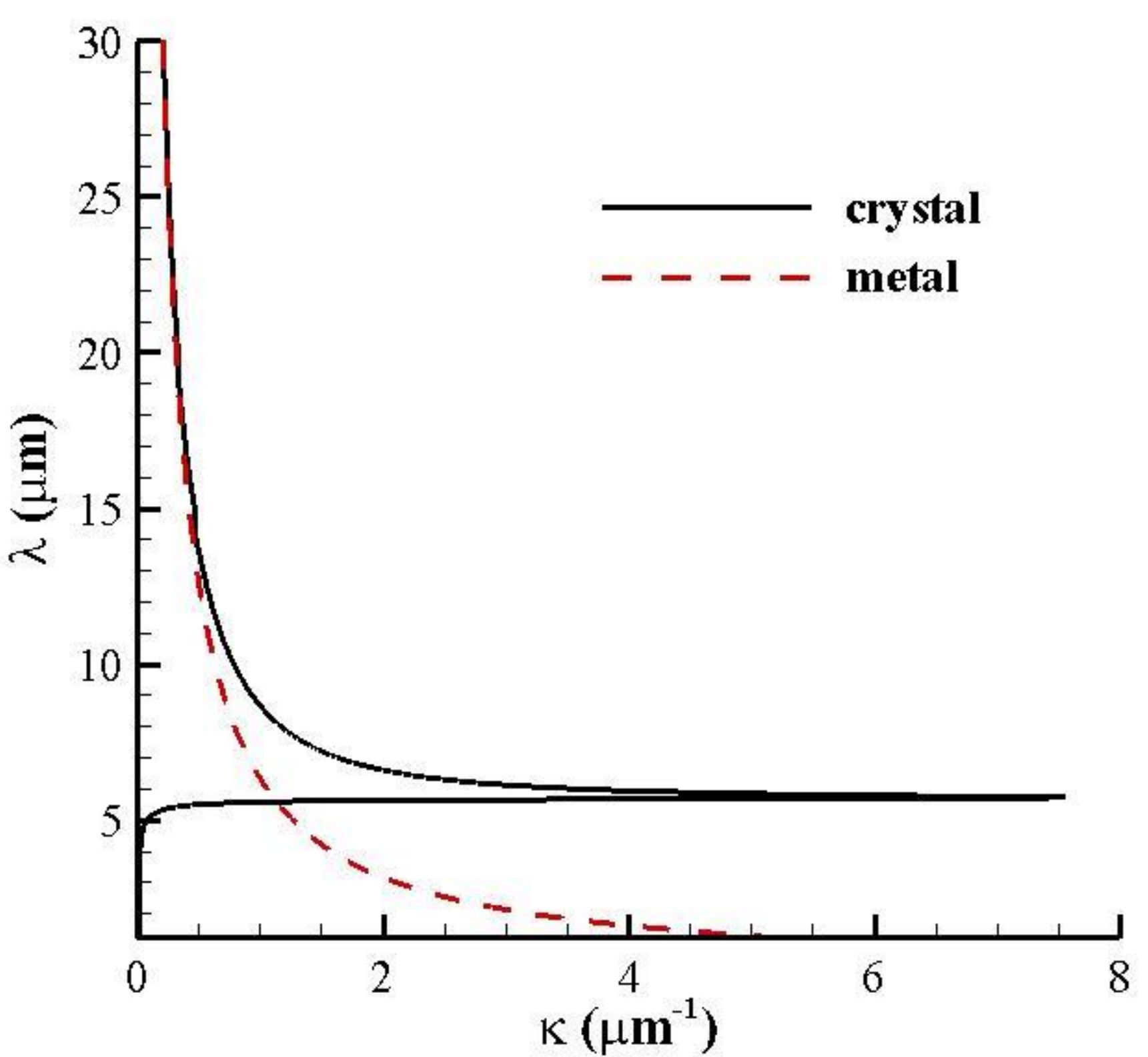}
\caption{Dispersion relation of surface modes of crystalline VO2. In its amorphous (metallic) state VO2 does not support any surface wave.\label{Fig:Dispersion}}
\end{figure}

Then, depending on the activated pattern, the far-field emission pattern of the source can be switched between two radically different shapes. The emissivity $\epsilon$ of these sources can be easily calculated from the reflection coefficient $R$ by using the Kirchoff's law (i.e. $\epsilon=1-R$). The result of this calculation is plotted in Fig. 4 in the $(\lambda,\theta)$ plane.  We see that when the source is in its crystalline state in all regions of space  the total thermal emission is very weak. This is consistent with the fact that the medium supports a surface wave so that the reflection coefficient is very large and therefore the absorption is weak. On the other hand, if we address some spatial regions so that a grating of VO2 emerges where some cells ar crystalline and others amorphous, then the emission pattern of the source becomes strongly directional and the magnitude of emission peaks is close to unity. 
\begin{figure}[Hhbt]
\includegraphics[scale=0.3]{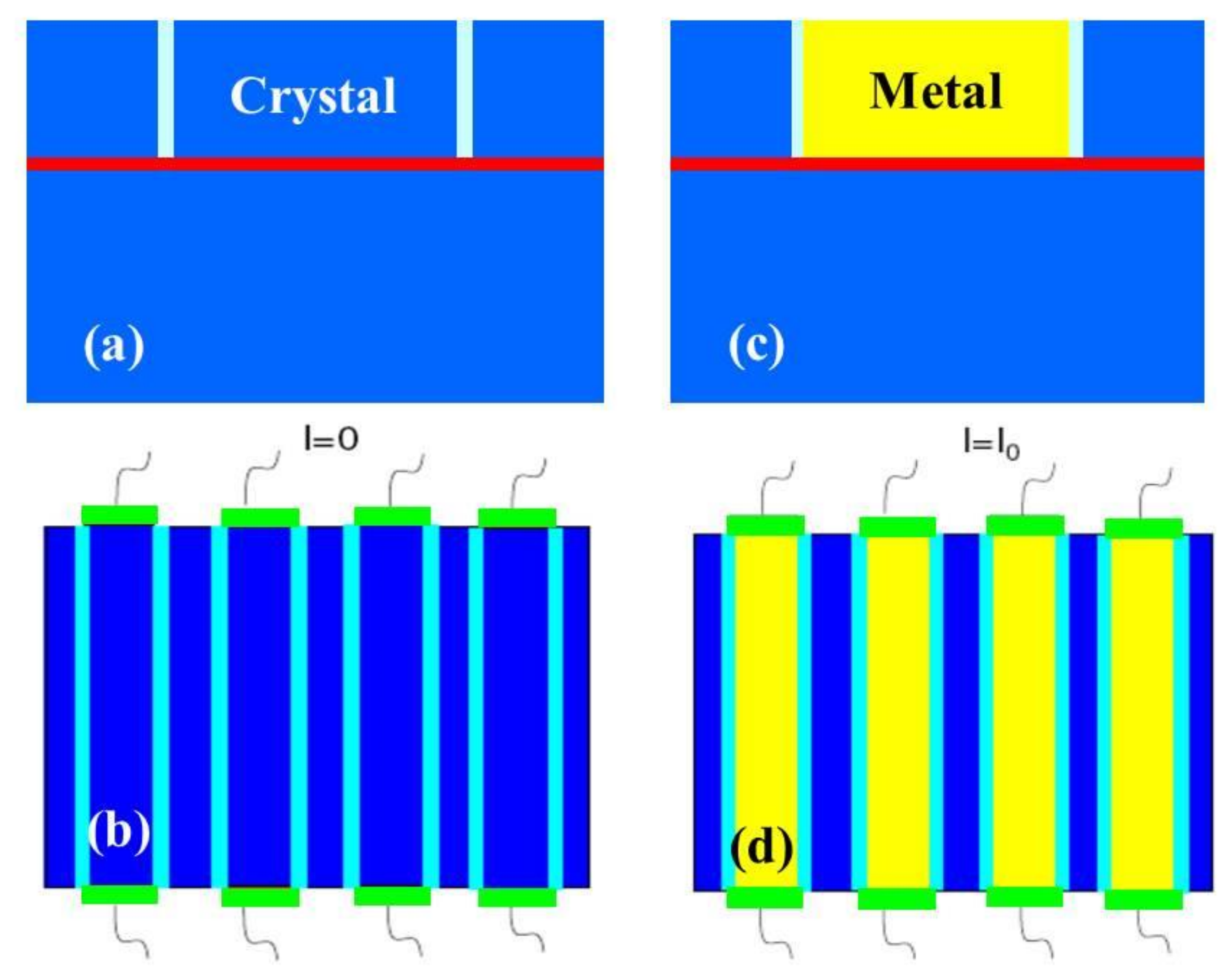}
\caption{Sketch of a thermal antenna based on phase transition materials. (a,b) the IMT material is in its crystalline phase and the source is non-radiating. (c,d) phase change is driven by an electrodes network (in green) to generate a half-metallic surface grating. A 20 nm thick thermally and electrically insulating layer of silica (in red) separates the grating from the substrate while grooves (in light blue) of 200 nm depth and 20 nm width filled by air separate all addressable VO2 cells.\label{Fig:structure}}
\end{figure}
From a practical point of view, to prevent the phase transition in unwanted regions, a thin thermally insulating layer of silica \cite{Palik} (typically $20 nm$ thick)  is introduced between the substratre and the region where the grating is generated by phase transition. The exact amount of heat that shall cross these boundaries depends on how critical it is to attain well separated phases, thus addressing the complex issue of hysteresis in relation with deposition technology.

\begin{figure}[Hhbt]
\includegraphics[scale=0.4,angle=-90] {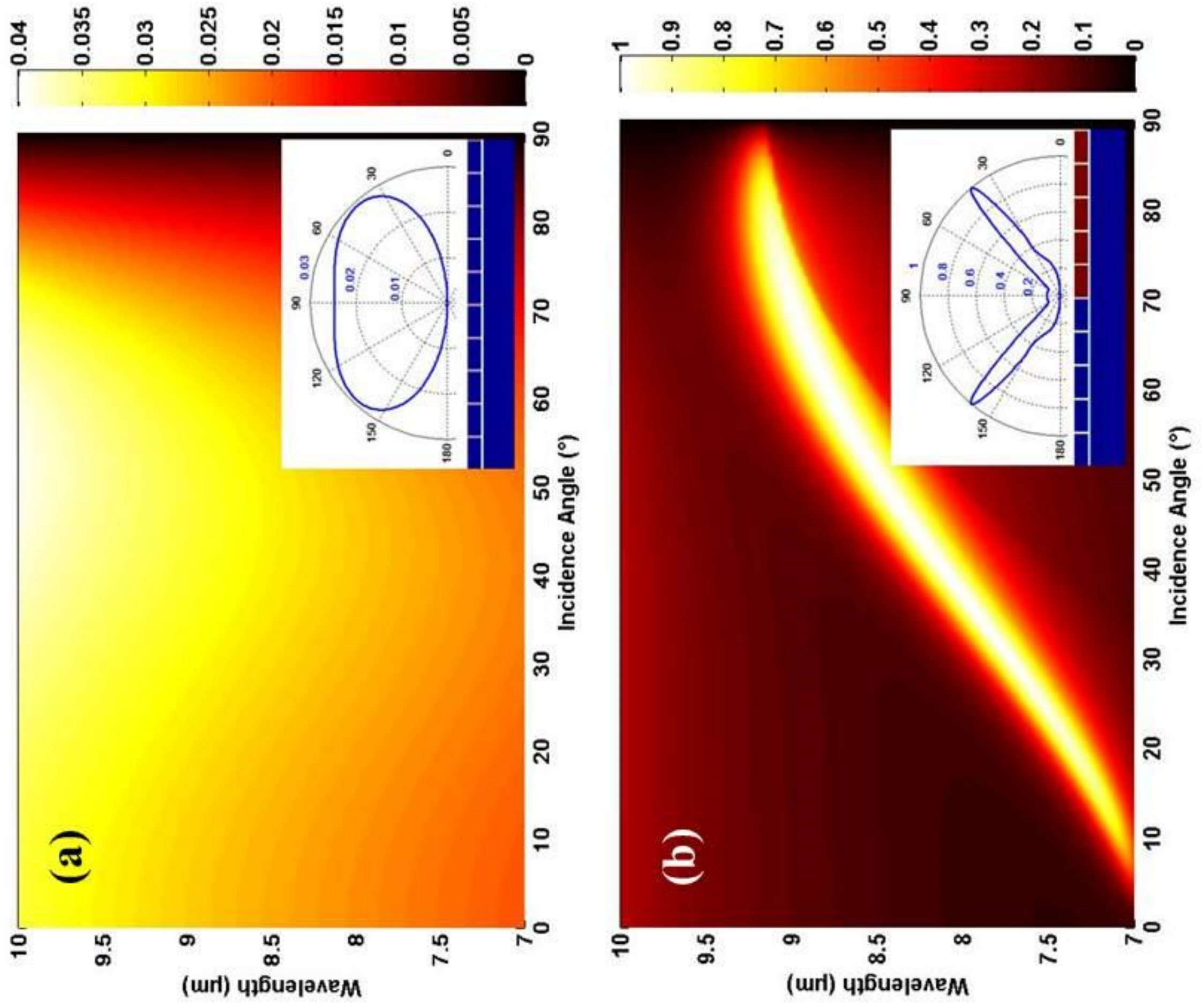}
\caption{TM-polarized emissivity of the VO2-source in (a) its purely crystalline (dark blue) non-emitting phase and (b) in its hybrid (crystalline-amorphous) phase where the amorphous phase forms a surface grating of period $P_1=4.5\mu m$ (10 unit cells of 450 nm length each). The inset shows the structure of one period of the source and its emission pattern at $8.5 \mu m$. \label{Fig:emissivity}}
\end{figure}

Finally, we show in Fig. 5 that we can also modulate the direction of emission of the antenna by addressing different VO2 cells so as to adequately form the geometric features of the surface grating. According to the theory of gratings, the direction of emission is given by the Bragg relation \cite{Kreiter}
\begin{equation}
 K_j=2\frac{\pi}{\lambda} sin \theta +2\frac{\pi}{P}j
\end{equation}
where $K_j$ is the parallel component of wavector diffracted in the $j^{th}$ order and P is the period of grating. This is similar to what is done for more classical spontaneous sources such as luminescent ones or light-emitting diodes, when they incorporate photonic crystals or gratings \cite{Henri}.

\begin{figure}[Hhbt]
\includegraphics[scale=0.31]{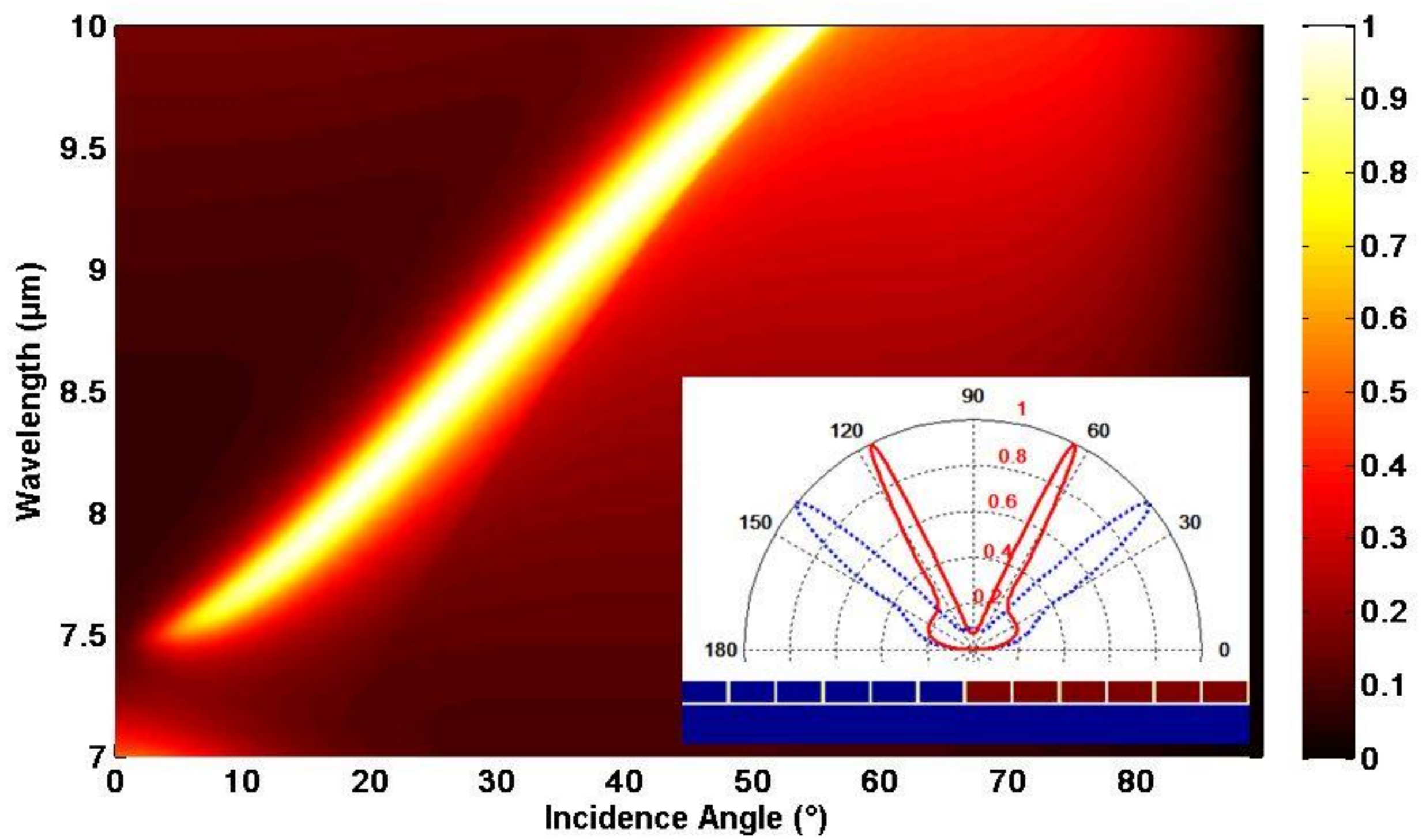}
\caption{Emissivity TM of the VO2-source in its antenna mode with a surface grating of period $P_2=5.4\mu m$ (12 unit cells of 450 nm length each. The inset shows the structural configation of the source and its emission pattern at $8.5 \mu m$. The blue dashed line represents the emision diagramm at the same wavelength when the grating period is $P_1$ as in Fig. 4. \label{Fig:modulation}}
\end{figure}

Of course these simulations address infinite periodic systems. From the $\Delta\theta~50  mrad$ typical angular width of the lobes on Figs.4-5, it is clear that the emitter size $D$ should verifiy $D>> P_{1,2}/\Delta\theta$ to avoid finite size effects and obtained the expected directionality. This means in practice that any size at or above the millimetre range is enough, but that for micro-electro-mechanical-Systems (MEMS) oriented applications with, say, $ D\sim100\mu m$, finite size effects, e.g., in the form of secondary lobes due to Fourier truncature should be considered in some detail. They would correspond to few tens of microwatt in terms of power dissipation by our proposed thermal antennae. Conversely, for macroscopic appliances with power rating in the range $10 mW\ldots100W$, finite size effects are of course irrelevant, only some complexities due to a hierarchy of patterned electrodes needed to spread current at the targeted locations may have to be considered \cite{Henri}.

In conclusion, we have introduced here a switchable thermal antenna based on the phase transition of a polaritonic IMT material, namely vanadium dioxide, a material currently incorporated in many applications and thus well mastered. To the best of our knowledge, there are no previous comparable proposals. To design switchable antenna that operates at lower or higher temperatures than the critical temperature of VO2 others IMT materials could be used.  We have shown that the thermal emission can be suppressed by the phase transition and that the direction of emission can be tailored by actively changing the geometric features of grating.

%
%

\begin{acknowledgments}
This work has been partially supported by the Agence Nationale de la Recherche through the
Source-TPV project ANR 2010 BLANC 0928 01. 
\end{acknowledgments}

\end{document}